\documentclass[twoside,a4paper,11pt]{sea10}
\usepackage{graphicx}
\usepackage{hyperref}
\usepackage{movie15}
\begin{document}
\pagenumbering{arabic}
\pagestyle{myheadings}
\thispagestyle{empty}
{\flushleft \includegraphics[width=\textwidth,bb=58 650 590 680]{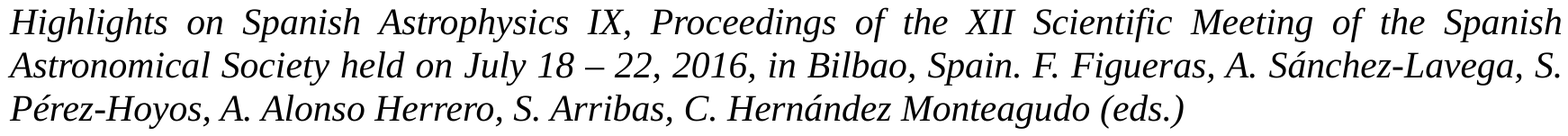}}
\vspace*{0.2cm}
\begin{flushleft}
{\bf {\LARGE
%
ESASky: The whole of space Astronomy at your fingertips 
%
}\\
\vspace*{1cm}
%
Bel\'en L\'opez Mart\'{\i}$^{1}$,
Bruno Mer\'{\i}n$^{1}$, Fabrizio Giordano$^{1}$, Deborah Baines$^{1}$, Elena Racero$^{1}$, Jes\'us Salgado$^{1}$, 
Mar\'{\i}a Henar Sarmiento$^{1}$, Ra\'ul Guti\'errez$^{1}$, Pilar de Teodoro$^{1}$, Juan Gonz\'alez$^{1}$, 
Juan Carlos Segovia$^{1}$, Sara Nieto$^{1}$, Henrik Norman$^{1}$,
and
Christophe Arviset$^{1}$
%
}\\
\vspace*{0.5cm}
%
$^{1}$
European Space Astronomy Centre (ESAC), Camino Bajo del Castillo s/n, Urb. Villafranca del Castillo, E-28692 Villanueva de la Ca\~nada, Spain \\
%
\end{flushleft}
%
\markboth{
ESASky: The whole of space Astronomy at your fingertips 
}{ 
%
B. L\'opez Mart\'{\i} \& ESASky Team
%
}
\thispagestyle{empty}
\vspace*{0.4cm}
\begin{minipage}[l]{0.09\textwidth}
\ 
\end{minipage}
\begin{minipage}[r]{0.9\textwidth}
\vspace{1cm}
\section*{Abstract}{\small
%
ESASky is a new science-driven discovery portal for all ESA astronomical missions that gives users worldwide a simplified access to high-level science-ready products from ESA and other data providers. The tool features a sky exploration interface and a single/multiple target interface, and it requires no prior knowledge of specific details of each mission. Users can explore the sky in multiple wavelengths, quickly see the data available for their targets, and retrieve the relevant products, with just a few clicks. The first version of the tool, released in May 2016, provides access to imaging data and a number of catalogues. Future releases will enable retrieval of spectroscopic data and will incorporate futures to allow time-domain exploration and the study of Solar System objects.
%
\normalsize}
\end{minipage}
%
%
%

\section{Introduction \label{sec:intro}}

The rapid growth in the volume of astronomical data obtained with telescopes and large infrastructures makes it compulsory for institutes and organisations to provide tools to their user communities for the efficient retrieval of those data, thus maximising the scientific return. Most current major space-based and ground-based facilities have already developed powerful, customisable archive interfaces allowing their users to carry out  very specific searches among a variety of data products, from raw to science-ready. However, these interfaces have usually been designed with the expert user in mind, and a relatively high level of knowledge of the particularities of each mission, telescope and/or instrument is required to efficiently exploit the available datasets.  As a consequence, users lacking this specific knowledge may refrain from using those data to complement their own research, or may not even know of the existence of the data and their potential usefulness for them.

On the other hand, the increasing importance of multi-wavelength studies in the current astrophysical research makes it necessary to develop tools that allow easy comparison and retrieval of multiple sorts of data, differing in wavelength coverage, resolution, type of observation (imaging, spectroscopy, etc.) and other relevant characteristics. The Virtual Observatory (VO) initiative already provides tools and standards enabling remote access to archives and comparison between data from different resources in a fast and efficient way. However, the use of most VO tools requires the previous download of these data before they can be inspected.

ESASky \cite{merin2015, merin2016} is a new multi-mission archive interface that has been developed at the ESAC Science Data Centre (ESDC) with the goal of enabling the search and retrieval of science data products from all ESA astronomical missions, as well as the comparison of datasets from the different missions, in a seamless and user-friendly manner. The application is conceived as a tool for exploration and discovery of the data kept in the different astronomical archives, and is thus targeted at a science user that does not need to be an expert in a particular mission or wavelength range. With ESASky, users can easily query and inspect the data available in several ESA scientific archives simultaneously, even if they have no previous knowledge about those missions; they can quickly compare the different datasets; and they can finally select and download the science-ready products of their interest, and only those products. The tool is completely web-based and requires no user registration.

The purpose of this contribution is to present the main features of the first release of ESASky, which was published in May 2016 and can be accessed at the following URL:

\begin{center}
\href{http://sky.esa.int/}{\tt http://sky.esa.int/}
\end{center}

\section{Main ESASky features \label{sec:features}}

The first version of ESASky addresses three use cases: i) exploration of multi-wavelength skies; ii) search and retrieval of data for single targets; and iii) search and retrieval of data for a target list.  We will now describe the elements and functionalities that enable the successful execution of these use cases. A video demoing all these functionalities can be found in the ESASky help pages.\footnote{\tt \href{http://www.cosmos.esa.int/web/esdc/esasky-video1}{http://www.cosmos.esa.int/web/esdc/esasky-video1}} For a complete technical description of the tool, please refer to \cite{merin2016}; a detailed description of the visualisation features within ESASky and their implementation is given in \cite{baines2016}.

At the moment of writing this contribution, the tool provides science-ready imaging data products and catalogue data for all ESA major astronomical missions, covering the whole wavelength domain from gamma-rays to radio: INTEGRAL, XMM-Newton, HST, Gaia, Hipparchos, Tycho, ISO, Herschel, and Planck. It also displays a large number of progressive all-sky maps from all ESA missions and from some external missions and surveys of interest for a broad part of the astronomical community. Full details on the scientific data content are given in the ESASky online help documentation.\footnote{\tt \href{http://www.cosmos.esa.int/web/esdc/esasky-help}{http://www.cosmos.esa.int/web/esdc/esasky-help}}

\subsection{Multi-wavelength skies exploration \label{sec:skies}}

ESASky enables exploration of the sky as seen by a large number of surveys and missions using so-called Hierarchical Progressive Surveys (HiPS) \cite{fernique2015}. A HiPS is an all-sky mapping of an astronomical dataset that uses a hierarchical tile and pixel structure of the sky based on the HEALPix tessellation of the sphere \cite{gorsky2005}, and that keeps the scientific properties of the original images. It is thus possible to visualise any part of the sky at progressively increasing resolution as the user zooms in, until the size of the native pixel from the images used to generate the HiPS is reached. An example HiPS is shown in Figure~\ref{fig:example}. 

HiPS can be visualised with a dedicated HiPS client. Currently ESASky uses AladinLite \cite{boch2014}, a lightweight version of the Aladin Sky Atlas \cite{bonnarel2000} running on browsers. Users can zoom in and out on a given region of the sky, pan with the mouse to move around  in the sky, or directly go to a target by entering its name or coordinates in a search box. The map itself can be visualised in either equatorial or galactic coordinates. It is possible to select and inspect any of the available skies, to change their color palette, and also to stack several of them to switch from one another (manually or using dedicated buttons). This way, it is pretty easy to visualise and move around the whole set of observations performed by a given mission or survey over the whole sky, and to compare observations from different missions and wavelength domains --all without having to download any data.

We generated HiPS by combining all the available images for the following ESA missions: INTEGRAL, XMM-Newton, EXOSAT, HST, ISO, AKARI, Herschel, and Planck. Detailed descriptions of all the HiPS, including information on the input data used to produce them, can be found in the ESASky documentation pages. In addition to them, ESASky also displays a number of skies from other missions and surveys that have been created by other data providers; a comprehensive list is also provided in the ESASky documentation. 

This multi-wavelength exploration functionality not only allows users to quickly check whether there are observations from a certain mission available for their object or region of interest (even from missions they did not know about!), but it also gives them the opportunity to decide, prior to any data download, whether the coverage, resolution or quality of those data adjusts to their scientific goals. Moreover, the simple comparison between observations by different missions and/or at different wavelength ranges can yield to very interesting scientific discoveries. It is important to keep in mind, though, that the HiPS are only intended for visualisation purposes; precise measurements of fluxes and positions of targets require the use of the actual science data.

\subsection{Visualisation and retrieval of imaging data \label{sec:data}}

In order to visualise the real data, the user has to go to his/her region of interest in the sky, adjust the size of the field of view if necessary, and open the data panel. Two histograms will be presented, one showing all the imaging data and the other all the catalogue sources found in that area of the sky. These histograms refresh automatically if the user changes the field of view or moves to another region of the sky. 

\begin{figure}[t!]
\center
\includegraphics[width=\textwidth,angle=0,clip=true]{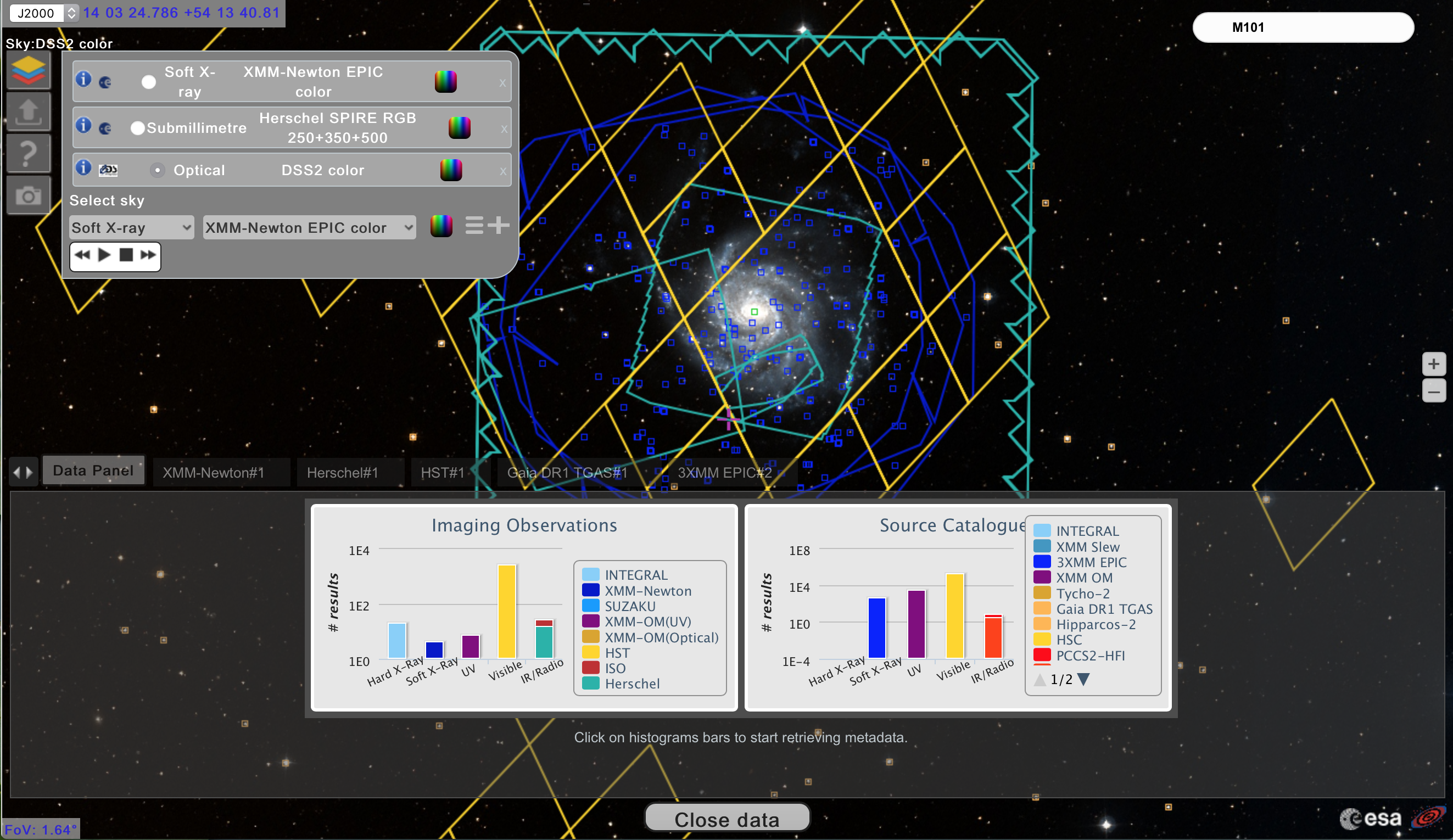} 
\caption{\label{fig:example} ESASky visualising galaxy M101 as seen in the optical by the Digital Sky Survey (DSS2). The skies menu is open in the top left and includes a number of stacked skies. The data panel showing the results histograms is open towards the bottom. Observation footprints are shown for XMM-Newton (blue) and Herschel (cyan), and a Multi-Order Coverage map indicating the approximate location of the numerous HST observations is shown in yellow. The sources from the 3XMM DR5 EPIC and the Tycho-Gaia Astrometric Solution catalogue (TGAS) within the field of view are displayed as small blue and orange squares, respectively.
}
\end{figure}

To inspect the data from a given mission, the user has to click on the corresponding histogram. A table will open in a tab within the results panel, providing a summary of the available observations, and at the same time, the footprints of the observations will be displayed on top of the sky map to show the exact region covered by each of them (Figure~\ref{fig:example} shows several examples). By selecting a table entry, the corresponding footprint is highlighted for a quick identification. It is also possible to visualise the observation postcard (preview) by clicking on the magnifying glass icon in the selected table row. 

Several sets of observations can be inspected simultaneously, allowing the user to compare their coverage and technical details. Once he/she has identified the observations that match his/her scientific goals, he/she can download all of them together in a tar file, getting the best-quality data products available for those particular observations. It is also possible to send the selected data products or the observations summary table to a VO application open in the user's desktop.

If the selected field of view is very large, the number of observations may get difficult to manage; in that case, the Multi-Order Coverage map (MOC) is displayed instead of the footprints of the individual observations (an example is also shown in Figure~\ref{fig:example}). A MOC is essentially a list of HEALPix cell numbers that provides a simple way to map regions of the sky into hierarchically grouped cells, which can be used to display the (approximate) coverage of a survey or catalogue \cite{fernique2014}. The use of MOCs allows quick large-scale comparisons between different missions and surveys (see e.g. \cite{lm2016}). To get the actual observational footprints is as easy as to decrease the size of the field of view and refresh the table. 

It must be stressed that ESASky is not an archive \emph{per se}, but just a gateway facilitating the access to the different mission archives: When the user selects some products for download, a query is sent to the corresponding science archive, which is the one providing the data. Hence, there is complete consistency between the data retrieved through ESASky and through the science archives themselves. As a matter of fact, each row in the observations summary table is linked to the corresponding entry in the mission science archive, which can be accessed by clicking on the observation ID. Because of this implementation, it is not necessary that the data themselves are hosted at ESAC: We can link to archives anywhere in the world. This is already done in the current ESASky version with the Suzaku imaging data, which can be downloaded through ESASky even if the files are physically stored at the mission archive at JAXA/ISAS in Japan.

\subsection{Visualisation and retrieval of catalogue data \label{sec:cats}}

Visualisation of catalogue data in ESASky is analogous to that of imaging data: By clicking on the corresponding histogram bar, a new tab opens displaying a table with all the catalogue sources within the selected field of view, and the positions of the sources in the sky are marked with small squares. By clicking on a table entry, the corresponding source location is highlighted. It is also possible to select a source by clicking on its position in the sky; in this case, the corresponding table row is highlighted, and a tooltip appears displaying basic information on that source. 

The catalogue table containing only the sources in the visualised field of view can be downloaded in CSV and in VOTable format, and can also be sent to a VO application.

\subsection{Working with target lists \label{sec:lists}}

A very common situation is when the user is not interested in a single target or region, but has instead a list of objects to be inspected. In this case, he/she can upload a target list with the objects' names or coordinates and (optionally) a description of each object, and navigate through this list to inspect each object individually. The histograms in the result panel will refresh automatically every time the application moves to a new object, but the different observation and catalogue tables will not, making it possible to leave open several tabs with the relevant data selected to be downloaded at the end of the process.

\section{Future plans \label{sec:plans}}

ESASky is in continuous development. New functionalities and data sets will be added in future releases to make the application more robust and complete, and to enable users to carry out other types of science. These are some of the features we are already working on:

\begin{itemize}

\item Visualisation and download of spectroscopic data from ESA missions. 

\item Time-domain exploration functionalities.

\item Functionalities for the visualisation and study of Solar System objects. Many missions include observations of planets and other objects in our Solar System, whose identification requires dedicated tools.

\item Functionality for the preparation of observations for live missions. Users will be able to display the field of view of the different instruments, to choose and modify their placement in the sky, and to compare them with observations already available.

\item Integration of data from third partners. Any institution wishing to publish their surveys or catalogues through our application can contact us via our Helpdesk.\footnote{\tt \href{https://support.cosmos.esa.int/esdc/index.php?/Tickets/Submit/}{https://support.cosmos.esa.int/esdc/index.php?/Tickets/Submit/}}

\end{itemize}

%
%
\small  
%
\section*{Acknowledgments}   
%
We acknowledge the excellent support from the expert science and technical staff at ESAC and CDS. In particular we acknowledge the following people: Mark Allen, Bruno Altieri, Ruben \'Alvarez, Guillaume Belanger, Thomas Boch, Tam\'as Budav\'ari, Javier Castellanos, Guido de Marchi, Xavier Dupac, Daniel Durand, Ken Ebisawa, Matthias Ehle, Pilar Esquej, Pierre Fernique, Pedro Garc\'{\i}a Lario, Krzysztof G\'orski, Jonas Haase, John Hoar, Peter Kretschmar, Erik Kuulkers, Ren\'e Laureijs, Daniel Lennon, Ignacio Le\'on, Nora Loiseau, Marcos L\'opez Caniego, Alejandro Lorca, Anthony Marston, Antonella Nota, William O'Mullane, I\~naki Ortiz de Landaluce, G\"oran Pilbratt, Andy Pollock, Roberto Prieto, Pedro Rodr\'{\i}guez, Michael Rosa, Miguel S\'anchez Portal, Maria Santos-Lleo, Norbert Schartel, Jan Tauber, Ivan Valtchanov and Eva Verdugo. 

ESASky makes use of Aladin Lite, developed at CDS, Strasbourg Observatory, France \cite{boch2014}. This work also benefited
from experience gained from projects supported by the Mission Operation Division of ESA's Directorate of Science.

%

%
\end{document}